\documentclass[sigconf]{acmart}

\usepackage[english]{babel}
\usepackage{blindtext}
\usepackage{tikz}
\usepackage{amsmath}
\usepackage{enumitem}
\usepackage{booktabs}
\usepackage{lineno}
\usepackage{xcolor}
\usepackage{bbding}
\usepackage{xurl}
\usepackage{hyperref}
\usepackage{subcaption}
\DeclareUrlCommand\code{%
  \urlstyle{tt}
}

\renewcommand\footnotetextcopyrightpermission[1]{} 
\setcopyright{none}

\acmDOI{}

\acmISBN{}

\acmPrice{}

\begin{document}
\title[SHIFT]{SHIFT: Exploring the Boundary of RDMA Network Fault Tolerance}

\author{\vspace{-2ex}Shengkai Lin$^{1,2}$, Kairui Zhou$^{1}$, Hongtao Zhang$^{2}$, Yibo Wu$^{2,3}$, Yi Pan$^{1,2}$, Yihan Yang$^{2}$, Qinwei Yang$^{1}$, Wei Zhang$^{4}$, Arvind Krishnamurthy$^{2}$, Shizhen Zhao$^{1}$}
\affiliation{%
  \institution{\vspace{-1.5ex}$^{1}$Shanghai Jiao Tong University \quad $^{2}$University of Washington \quad $^{3}$University of Wisconsin-Madison \quad $^{4}$University of Connecticut\vspace{-2ex}}}

\renewcommand{\shortauthors}{Lin, et al.}

\begin{abstract}
Under gang scheduling for large-scale distributed large language model (LLM) training, a single network anomaly can stall or abort an entire job. Current network fault tolerance mechanisms typically adopt a ``fallback and bypass'' approach within the switching fabric and at the access layer, tolerating in-network and access-layer failures.

We explore whether RDMA fault tolerance can be extended to the cross-NIC level by failing over traffic to intra-host backup NICs. For the first time, we prove a fundamental Trilemma: it is impossible to have Cross-NIC RDMA failover that simultaneously preserves Exactly-Once Execution, Receiver-NIC Opacity, and a Zero-Copy datapath.

Fortunately, we observe that dominant training frameworks (e.g., NCCL) rely on idempotent bulk transfers that tolerate relaxed memory ordering, as long as notification ordering is preserved. Leveraging this insight, we present SHIFT, a user-space RDMA layer that provides cross-NIC fault tolerance while preserving correct memory semantics.
We implement SHIFT in \texttt{rdma-core} and evaluate it with PyTorch distributed training. Results show that SHIFT incurs negligible overhead during normal operation and successfully masks fatal NIC failures and link anomalies, allowing training to continue without costly restarts.
\end{abstract}

\maketitle

\section{Introduction}
\label{sec:intro}

Large language models (LLMs)~\cite{touvron2023llama, gpt3, llama3} continue to scale rapidly. State-of-the-art models already require tens of thousands of GPUs interconnected by high-speed fabrics, such as Remote Direct Memory Access (RDMA) networks, to sustain high-throughput training~\cite{zhang2022opt, megascale, aegis}.

As LLMs scale and rail-optimized networks increase the number of optical interconnects, network anomalies become more frequent. Production data from Alibaba indicates that network-related issues account for 15.8\% of all production failures~\cite{aegis}. Under gang scheduling for distributed training, a single anomaly can abort the entire job, wasting computational resources and losing training progress~\cite{character, hpn}.

\begin{figure}[t]
    \centering
    \includegraphics[width=0.4\textwidth]{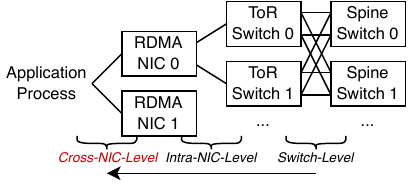}
    \setlength{\abovecaptionskip}{0pt}
    \setlength{\belowcaptionskip}{-20pt}
    \caption{
    \textit{Levels of network fault tolerance. Existing deployments support ``fallback and bypass'' within the switching fabric and at the access layer. SHIFT extends this capability to the cross-NIC level.}
    }
    \label{fig:faulttolerance}
\end{figure}

Given the fragility of training workloads, network fault tolerance is increasingly critical.
The prevailing paradigm in this domain is ``fallback and bypass'': reroute around failures and continue transmission.
As depicted in Figure~\ref{fig:faulttolerance}, failures within the switching fabric are typically handled by in-network rerouting, while failures at the access layer (e.g., access links and ToR switches) can be tolerated through dual-port NICs or dual-ToR designs~\cite{hpn,astral}.

This status quo raises a fundamental question: with the widely deployed RDMA Reliable Connection (RC) transport (exactly-once delivery), commodity RDMA hardware, and zero-copy data paths, \emph{have we reached the boundary of RDMA network fault tolerance?}
Can we push beyond current solutions and hide NIC-level anomalies from training jobs?

We argue that the answer is both \textit{negative and positive}.
In this paper, we extend ``fallback and bypass'' to the cross-NIC level (Figure~\ref{fig:faulttolerance}).
Our objective is to enable a system that, upon detecting an RNIC anomaly or an anomaly that is not masked by in-network or access-layer mechanisms, retransmits in-flight traffic and seamlessly falls back to another intra-host NIC.
Such a mechanism would bypass NIC-to-NIC failures and shield the application from endpoint fragility.

However, we identify a fundamental constraint: the widely used RDMA RC protocol provides a delivery guarantee that is \textit{too strong} for transparent cross-NIC fault tolerance.
We prove that, with commodity RNICs and application-agnostic requirements, seamless cross-NIC fault tolerance is \emph{impossible} without violating RC's ``exactly-once'' guarantee.
This constraint blocks support for workloads that rely on strict ordering semantics, such as atomic operations (\S\ref{sec:insight_boundary}).

Fortunately, we also observe that dominant communication patterns in training frameworks rely on idempotent bulk transfer that tolerate relaxed memory ordering.
This observation provides the flexibility needed to circumvent RC's rigidity, making it feasible to tolerate NIC-to-NIC RDMA failures for training workloads. (\S\ref{sec:opportunity})

We propose \textbf{SHIFT}, which exploits intra-host NICs as mutual backups.
Upon detecting a network anomaly, SHIFT retransmits affected RDMA traffic and redirects it through a backup NIC on the same host, bypassing the failure.
Once the default NIC recovers, SHIFT switches traffic back.
Although communication through a backup NIC can be constrained by PCIe bandwidth and can interfere with other co-located traffic, tolerating anomalies can still be advantageous by allowing training to continue \textit{until the next checkpoint completes or network connectivity recovers}.

This design hides transient anomalies from training workloads, and it enables fatal failures to be handled without immediate crash-stop, minimizing training progress loss.
SHIFT remains orthogonal to application-layer solutions.

Our contributions are summarized as follows:

\begin{enumerate}[noitemsep,parsep=0pt,partopsep=0pt,topsep=2pt,leftmargin=15pt]
\item We prove that transparent cross-NIC fault tolerance cannot preserve RC semantics (exactly-once) under commodity RNICs and a zero-copy data path. Through an analysis of NCCL~\cite{nccl}, NVSHMEM~\cite{nvshmem}, and MSCCL++~\cite{msccl++}, we show that dominant training traffic tolerates weaker semantics, enabling effective cross-NIC fault tolerance. (\S\ref{sec:overview})
\item We exploit WR-level retransmission and propose SHIFT, which introduces three key mechanisms: (1) copying WQEs, (2) CQ event-based two-way handshakes, and (3) WR execution fences. We implement SHIFT within the RDMA userspace library~\cite{rdma-core}. (\S\ref{sec:design})
\item We evaluate SHIFT's bandwidth, latency, and overhead using microbenchmarks and assess its impact on real-world distributed training. Our results confirm that SHIFT delivers robust RDMA resiliency with negligible performance overhead. (\S\ref{sec:evaluation})
\end{enumerate}

\emph{This work does not raise any ethical issues.}

\section{Background and Motivation}
\label{sec:background}
This section provides an overview of large-scale AI training and its failure landscape (\S\ref{sec:landscape}), reviews existing network-layer fault tolerance mechanisms (\S\ref{sec:net_ft}), and introduces the basics of RDMA (\S\ref{sec:rdma_basics}).

\subsection{Failures in Large-Scale AI Training}
\label{sec:landscape}
Large language models (LLMs) have scaled to unprecedented sizes. For instance, Llama 3~\cite{touvron2023llama, llama3}, GPT-3~\cite{gpt3}, and GPT-4 contain hundreds of billions of parameters. Training these models requires massive clusters with tens of thousands of GPUs~\cite{hpn, insightsdeepseek}, typically interconnected by high-performance RDMA networks.

However, the synchronous nature of distributed training renders it inherently fragile. A standard training iteration involves collective communication where all workers must synchronize. Consequently, a failure in any single component (a GPU, NIC, or cable) stalls the entire training task. This phenomenon is often referred to as the "straggler problem" and may result in a complete task crash~\cite{character, hpn}.

To mitigate failure impact, current training frameworks propose two types of primary mechanisms:
\textbf{Checkpointing-restart}~\cite{deepfreeze, checknrun, checkfreq, gemini, ByteCheckpoint} serves as the standard approach by periodically saving model states to persistent storage. Upon failure, training resumes from the last stored checkpoint.
\textbf{Runtime resilience} techniques~\cite{bamboo, oobleck, trainmover, recycle} offer a more agile alternative. They dynamically replace failed workers or reconfigure the pipeline to exclude faulty nodes, thereby reducing downtime compared to full restarts.

Despite these efforts, these application-layer solutions remain suboptimal.
Checkpointing inherently incurs "progress loss": all computation performed since the last checkpoint is discarded upon failure. While increasing checkpoint frequency reduces this loss, it introduces significant I/O and network overhead.
Besides, runtime resilience mechanisms often require specific parallelism strategies (e.g., pipeline parallelism) or redundant resources (e.g., hot standbys), which limits their universality and compatibility.

\textbf{\emph{Takeaway} \#1}
Distributed training is highly sensitive to hardware anomalies. Existing application-layer fault tolerance mechanisms alleviate the impact but suffer from progress loss, resource overhead, or limited universality.

\subsection{Tolerate Network Anomalies}
\label{sec:net_ft}
In modern AI training clusters, network anomalies have emerged as a significant source of hardware failures due to two primary factors:

\begin{enumerate}[noitemsep,parsep=1pt,partopsep=0pt, topsep=0pt, leftmargin=15pt]
\item \textbf{Scale:} As cluster sizes expand, the absolute frequency of network anomalies increases commensurately, even with constant per-device failure rates.
\item \textbf{Optical Interconnects:} The adoption of rail-optimized networks~\cite{ncclall2all, superpod, hpn} necessitates optical fibers for NIC-to-switch connections. Optical transceivers exhibit  higher failure rates than copper cables. 
\end{enumerate}

Production data corroborates this trend: Alibaba attributes \textbf{15.8\%} of \emph{all training failures} to network issues (9.1\% NICs, 6.7\% optics)~\cite{aegis}, while Azure reports that \textbf{8.3\%} of failures in InfiniBand clusters are network-related~\cite{anubis}. Tencent further reports that NIC errors account for \textbf{15\%} of \emph{network anomalies}, while switches and fiber account for an additional \textbf{30\%} of network anomalies~\cite{astral}. These anomalies manifest as either fatal failures (e.g., device breakdown) or transient failures (e.g., interface flapping~\cite{megascale}), both of which stall or disrupt training progress.

Given the limitations of application-layer fault tolerance solutions, enhancing fault tolerance at the network layer is critical. The objective is to \emph{mask network failures} from the application, typically adhering to the "fallback and bypass" paradigm, as shown in Figure~\ref{fig:faulttolerance}.

\textbf{In-network Rerouting.} Failures within the network fabric (e.g., switch-to-switch links) are typically handled via in-network rerouting. Data center networks typically provide multiple physical links and paths (e.g., ECMP), allowing traffic to seamlessly bypass faulty devices without application intervention.

\textbf{Access-Layer Redundancy.} Failures at the access layer (between the ToR switch and the RNIC) are also addressed through redundancy:
RoCE LAG (Link Aggregation)~\cite{rocelag} bonds the two ports of the same NIC into a single logical interface to tolerate single cable or port failures.
Dual-ToR architectures~\cite{hpn} further connect a dual-port RNIC to two different ToR switches to tolerate ToR switch failures.

However, these mechanisms face two limitations:
(1) They fail to tolerate failures of RNIC itself, which constitute a significant portion of production incidents (9.1\% in Alibaba's cluster~\cite{aegis}).
(2) They impose extra hardware dependencies (e.g., dual-port NICs) and network configurations, restricting deployment flexibility.

\textbf{\emph{Takeaway} \#2}
While in-network rerouting handles in-network failures, existing access-layer solutions leave RNIC failures unhandled and lack deployment flexibility. 
This prompts the question: Given the common multi-NIC architecture of modern GPU servers, can we leverage it and achieve \emph{cross-NIC fault tolerance}? We investigate this in \S\ref{sec:overview}.

\subsection{RDMA Basics}
\label{sec:rdma_basics}
We briefly review the basics of RDMA relevant to our design.
RDMA serves as the standard for high-throughput, low-latency communication in AI clusters. 
The RDMA userspace library~\cite{rdma-core} exposes the verbs API, managing connections via Queue Pairs (QPs). Each QP comprises a Send Queue (SQ) and a Receive Queue (RQ), associated with a Completion Queue (CQ). 

QPs support multiple transport modes, including Reliable Connection (RC), Unreliable Connection (UC), and Unreliable Datagram (UD)~\cite{transportmodes}. This work focuses on RC, the predominant transport for training workloads, which guarantees reliable delivery.

RDMA operations fall into two categories: one-sided (e.g., \textit{write}) and two-sided (e.g., \textit{send}, \textit{write with immediate}\footnote{\textit{Write with immediate} transfers data within work requests and consumes a receive work request from the receive queue.}). 
To initiate an operation, the CPU posts a Work Request (WR) to the SQ via \code{post_send}. The driver converts this WR into a Work Queue Element (WQE) in registered memory and rings the RNIC's doorbell register to trigger processing.
For \textit{signaled} WRs, the application retrieves a Work Completion (WC) from the CQ via \code{poll_cq} upon completion. Unsignaled WRs generate no notification unless an error occurs.

For two-sided operations, the receiver must pre-post a receive WR to the RQ using \code{post_recv}. Due to this requisite CPU involvement, two-sided operations incur higher overhead and are typically used for control messages.

In RC transport, the receiver sends an acknowledgment (ACK) only after successfully writing the data to memory. The sender RNIC generates a successful WC upon receiving this ACK~\cite{qptype}.
If packet loss persists beyond the retransmission limit, the sender RNIC reports an error WC and transitions the QP to an error state.

\textit{Refer to Appendix~\ref{app:rdmadetail} for further technical details on RDMA.}

\section{The Boundary of RDMA Network Fault Tolerance}
\label{sec:overview}

\textbf{\textit{Cross-NIC Fault Tolerance.}}
In this paper, we propose SHIFT, a mechanism that pushes the fallback and bypass paradigm further to the cross-NIC level, therefore tolerating NIC-to-NIC failures.
The core insight is that modern GPU servers are typically equipped with multiple RNICs that are connected to GPUs via the PCIe bus. It is therefore feasible for intra-host RNICs to \textit{serve as backups for one another}. 
When a network anomaly occurs (e.g., NIC failure or link failure), SHIFT seamlessly migrates ongoing traffic to a backup RNIC (\textbf{passive switching}). Once the default RNIC recovers (e.g., after an interface flapping), traffic is seamlessly migrated back to restore the original setting (\textbf{active switching}).
This allows training to continue uninterrupted, providing a more resilient RDMA network.

Although communication through a backup RNIC may be constrained by PCIe bandwidth and may interfere with other traffic, it remains advantageous to tolerate network anomalies and allow training to continue \textit{until checkpointing or network connectivity recovers}. Applications also have the flexibility to just complete training with the fallback setting.

However, this intuition faces theoretical hurdles. We demonstrate that while maintaining strict RC semantics during cross-NIC failover is impossible (\S\ref{sec:insight_boundary}), the relaxed consistency requirements of training workloads offer an opportunity for SHIFT to achieve practical fault tolerance (\S\ref{sec:opportunity}).

\subsection{RDMA Failover Trilemma}
\label{sec:insight_boundary}

We discover the following RDMA Failover Trilemma: no mechanism can simultaneously achieve \textbf{Exactly-Once}, \textbf{Receiver NIC state Opacity}, and \textbf{Zero-Copy}. All proofs are mechanically verified in Rocq 9.0 ($\sim$3,500 lines)~\cite{rocq}; formal details are in Appendix~\ref{app:proofs}.

We first explain the three requirements for RDMA failover.

(1) \textbf{Exactly-Once Execution}: RDMA guarantees \textbf{exactly-once} delivery and processing of messages when using RC Queue Pairs. Exactly-once is two-fold: 1) \emph{Liveness} ensures that every operation eventually executes; 2) \emph{Safety} enforces that each operation executes at most once.

(2) \textbf{\emph{Receiver-NIC-Opacity}}: we assume no knowledge of the receiver NIC's internal states. In practice, RNIC's internal states (e.g., receiver $psn$) can be volatile or even inaccessible to receiver software (e.g., the receiver NIC can fail right after transferring the data).

(3) \textbf{\emph{Zero-Copy Datapath}}: the data should be transferred at most once. Additional copy wastes CPU cycles.

\textbf{The Safety-Liveness Conflict.} Theorem~\ref{thm:impossibility_main} captures a fundamental conflict between liveness and safety when Receiver-NIC-Opacity and Zero-Copy Datapath are enforced.

\begin{lemma}[Indistinguishability]
\label{lem:impossibility}
There exist two traces indistinguishable yet conflicting without receiver NIC state (See Appendix~\ref{app:proof_indistinguishability}).
\end{lemma}

Consider two traces $\mathcal{T}_1$, $\mathcal{T}_2$: (1) \textbf{Packet Lost ($\mathcal{T}_1$)}: The request was lost before NIC failure. The operation was never executed. 
(2)\textbf{ACK Lost ($\mathcal{T}_2$)}: The operation was executed before NIC failure, but the ACK was lost. 

Note that $\mathcal{T}_1$ and $\mathcal{T}_2$ are identical in the sender's view: $[\texttt{Sent}, \texttt{Timeout}]$. So, any decision function without receiver state information will return the same result. Then, if the sender decides not to resend the packet, liveness may be violated; if the sender decides to resend the packet after failover, this packet may be received twice.

\begin{lemma}[Non-Idempotency]
\label{lem:non-idempotency}
RDMA operations are non-idempotent\footnote{Idempotency means that an operation can be applied multiple times without changing the result beyond the initial application.} if we preserve zero-copy (See Appendix~\ref{app:proof_non_idempotent}).
\end{lemma}

RNICs can drop retransmitted packets to ensure exactly-once execution. However, upon NIC failure, the receiver's state may get lost and thus the backup NIC may accept a packet that is already accepted by the previous NIC. One may introduce a staging buffer to track the receiver status in memory. But this violates the zero-copy requirement. With zero-copy, some RDMA operations may be executed more than once. Unfortunately, there exist RDMA operations are non-idempotent:
\begin{itemize}[noitemsep,parsep=0pt,partopsep=0pt,topsep=2pt,leftmargin=12pt]
\item \textbf{Two-sided Operation}: A two-sided RDMA operation would consume a receiver's receive work request.
\item \textbf{RDMA Atomics (FADD/CAS)}: Retrying $\texttt{FADD}(a, \delta)$ results in adding $2\delta$; retrying a $\texttt{CAS}(0 \to 1)$ after a concurrent reset ($1 \to 0$) causes a double-success (ABA problem), violating linearizability.
\item \textbf{Complex RDMA Write}: Applications that pack data and notification flags in a single write (e.g., \textbf{NCCL LL/LL128}) or use a dedicated signal write (``Data + Signal'') are non-idempotent. A full retry may cause the receiver to overwrite memory that is already used by the application.
\end{itemize}

\begin{theorem}[Impossibility]
\label{thm:impossibility_main}
A receiver-NIC opaque, zero-copy cross-NIC RDMA failover mechanism cannot simultaneously preserve safety and liveness for systems with non-idempotent operations.
\end{theorem}

Since RDMA provides Atomic Operations, one might hypothesize that this ambiguity could be resolved through clever bookkeeping strategies or a sophisticated verification protocol from the sender. We prove that this is fundamentally impossible through a distributed system perspective:

\begin{theorem}[Consensus Hierarchy Barrier]
\label{thm:consensus}
Solving the retransmission ambiguity under failure without receiver NIC state is equivalent to solving 2-process consensus with only read/write primitives (See Appendix~\ref{app:proof_consensus}).
\end{theorem}

Intuitively, the failover mechanism must implement a ``First-Writer-Wins'' policy with RDMA operations. Such a policy is equivalent to a Sticky Register (with Consensus Number 2), which is impossible given that our failure model is essentially a non-responsive omission failure~\cite{jayanti1998fault}.

\textbf{\emph{Takeaway} \#3}
These impossibility results define the RDMA Failover Trilemma: a fault tolerance mechanism cannot simultaneously satisfy Exactly-Once Execution, Receiver-NIC Opacity, and a Zero-Copy Datapath. Given this trilemma, we must relax one constraint for the design of failover strategies:

\begin{itemize}[noitemsep,parsep=0pt,partopsep=0pt,topsep=2pt,leftmargin=12pt]
\item \textbf{Zero-Copy} is non-negotiable. Reintroducing the CPU into the datapath for buffering or logging fundamentally nullifies the performance benefits of RDMA.
\item \textbf{The receiver-NIC} is a hardware reality and impossible to overcome because: 1) Fixed Logic: Commodity hardware prevents injecting custom logic into the offloaded datapath. 2) Vendor Lock-in: Internal states are proprietary and lack standardized APIs. 3) Hard Failures: In cases of total hardware failure (which we target), the NIC state is not just opaque—it is destroyed.
\item \textbf{Exactly Once Execution}: we find that this semantic may be too strong for AI workloads, which opens an opportunity for cross-NIC failover. 
\end{itemize}

\begin{table*}[h]
\centering
\footnotesize
\begin{tabular}{p{2.5cm} p{4cm} p{10cm}}
\toprule
\textbf{Library \& Protocol} & \textbf{RDMA Ops (Data / Notify)} & \textbf{Reliable Requirements} \\
\midrule

\textbf{NCCL (Simple)}~\cite{demystifying}
& 
\textbf{Data:} \texttt{RDMA Write} \newline
\textbf{Notify:} \texttt{RDMA Write\_Imm}
& 
The receiver only accesses the received data after getting the following notification's WC. Require ordering between \texttt{RDMA Write} and the following \texttt{RDMA Write\_Imm}.
\\ 
\midrule

\textbf{NVSHMEM}~\cite{nvshmem}
& 
\textbf{Data:} \texttt{RDMA Write} \newline
\textbf{Notify:} \texttt{RDMA Atomic}
& 
The receiver only accesses the received data after polling the signal variable updated by the atomic operation. In addition to ordering, this requires the atomic operation to execute exactly once.
\\ 
\midrule

\textbf{MSCCL++}~\cite{msccl++}
& 
\textbf{Data:} \texttt{RDMA Write} \newline
\textbf{Notify:} \texttt{RDMA Atomic}
& 
Same requirements as NVSHMEM. \\
\midrule

\textbf{NCCL (LL/LL128)}~\cite{demystifying} & 
\textbf{Data + Notify:} \texttt{RDMA Write} \newline
(Packed: 4B Data + 4B Flag or 120B Data + 8B Flag) 
& 
Since both data and flags are transferred via pure writes, strict byte-level ordering and an exactly-once guarantee are required. \\ 
\bottomrule
\end{tabular}
    \setlength{\abovecaptionskip}{0pt}
\caption{Comparison of common communication libraries' protocols in training workloads.}
\label{tab:rdma-resilience}
\end{table*}

\subsection{Opportunity}
\label{sec:opportunity}
We investigated common communication libraries, including NCCL~\cite{nccl, demystifying}, NVSHMEM~\cite{nvshmem}, and MSCCL++~\cite{msccl++}, as summarized in Table~\ref{tab:rdma-resilience}. The related RDMA operations are summarized below:
\begin{enumerate}[noitemsep,parsep=0pt,partopsep=0pt,topsep=2pt,leftmargin=15pt]
    \item NCCL (Simple), \emph{the most prevalent} protocol, transmits bulk data via multiple RDMA writes followed by a \code{write_with_imm} that serves as a completion notification to the receiver. The receiver accesses the data buffer only after getting this notification.
    \item NVSHMEM and MSCCL++ transfer bulk data with RDMA writes but use atomic operations for notification~\cite{nvshmem,msccl++}.
    \item NCCL (LL/LL128) are low-latency protocols that pack both data and notification flags into a single RDMA write. However, LL only utilizes 25--50\% of peak bandwidth, which conflicts with high-bandwidth training requirements~\cite{hpn,metascale}; LL128 requires specialized hardware capabilities (byte-level ordering)~\cite{demystifying}. As a result, these protocols are rarely used for training workloads.
\end{enumerate}

\noindent\textbf{Opportunity:} These observations yield the key opportunity for cross-NIC fault tolerance:

For these common protocols, \emph{bulk data transfers are effectively idempotent at the application level.} However, applications still require ordering between the data transfer and its notification.

Therefore, it is possible to provide practical cross-NIC fault tolerance as long as it preserves the ordering of notifications.

\noindent\textbf{Design intuition:} For NCCL (Simple), the design intuition for cross-NIC fallback is:
\begin{itemize}[noitemsep,parsep=0pt,partopsep=0pt,topsep=2pt,leftmargin=12pt]
    \item The last successfully completed receive WR (notification) indicates that the data of WRs before it has been completely written to receiver memory. In contrast, data of WRs after it has not yet been consumed and remains safe to retransmit.
    \item Since the last acknowledged send packet is opaque to software, SHIFT can only identify sender-side progress via the last successfully completed send WR. Therefore, SHIFT retransmits starting from the first failed send WR after the last successfully completed receive WR.
\end{itemize}

On the other hand, for NVSHMEM and MSCCL++, which utilize one-sided atomic operations for notification, SHIFT cannot guarantee ordering between data transfer and notification, as proved in \S\ref{sec:insight_boundary}. To prevent data corruption, SHIFT must prohibit retransmission \textbf{IF} an atomic WR is in flight and instead return an application error.
SHIFT does not support NCCL LL/LL128 and must be disabled by network operators when these protocols are used.

\textbf{\emph{Takeaway} \#4}
Exploiting the relaxed RC requirements of training workloads, SHIFT provides bounded cross-NIC fault tolerance: it preserves notification ordering while allowing retransmission of idempotent bulk data transfers.
SHIFT is able to handle the majority of training traffic without compromising memory consistency.

\section{SHIFT Design}
\label{sec:design}
\begin{figure}[t]
    \centering
    \includegraphics[width=0.37\textwidth]{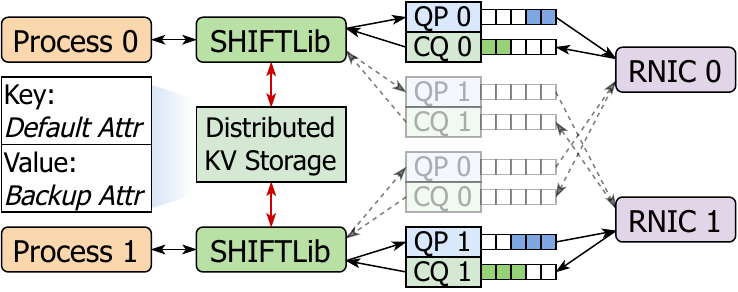}
    \setlength{\abovecaptionskip}{0pt}
    \setlength{\belowcaptionskip}{-10pt}
    \caption{
    \textit{Architecture of SHIFT. The blue boxes and the green boxes denote the resources managed by the application and SHIFTLib, respectively. The black arrows and the red arrows denote the data flows and the control flows, respectively.}
    }
    \label{fig:architecture}
\end{figure}

\subsection{Design Overview}
\label{sec:designoverview}
SHIFT achieves cross-NIC fault tolerance by retransmitting from the first failed send WR submitted after the last successfully completed receive WR. SHIFT adopts three key designs for both passive and active switching.

\textbf{1. HOW to resubmit under passive switching.} Achieving seamless RDMA traffic switching under passive switching requires resubmitting WRs that have been submitted but not yet completed to the backup RNIC. To keep this process application-agnostic, a naive approach is to add a WR buffer to each QP and locally store all submitted but uncompleted WRs. Passive switching can then be achieved by resubmitting WRs from this buffer~\cite{luberdma}. However, such buffering introduces CPU overhead and data path latency. Furthermore, the buffer's memory footprint grows linearly with the number of QPs.

\textit{\underline{Solution: Copying inherent WQEs.}} 
As mentioned in \S\ref{sec:rdma_basics}, WRs are converted into WQEs and stored in WQs, which reside in host memory. Therefore, WQEs remain valid in memory despite network failures. SHIFT can thus recover WQEs from the default QP's WQ, modify a few attributes, and resubmit them to the backup QP. See details in \S\ref{sec:faulttolerance}.

\textbf{2. WHEN and WHAT to resubmit under passive switching.} 
Two-sided operations require that receive WQEs be posted in the receive queue before the matching send WQEs arrive; otherwise, the send WQEs trigger a Receive-not-Ready (RNR) error. 
Furthermore, as analyzed in \S\ref{sec:opportunity}, SHIFT needs receive-side progress to determine which WQEs should be retransmitted. These requirements necessitate a synchronization mechanism between the sender and receiver.

\textit{\underline{Solution: CQ event-based 2-way handshake.}} 
To record send and receive progress, SHIFT tracks the number of completed send and receive WQEs locally. To minimize CPU overhead, SHIFT leverages RDMA CQ events, which trigger user-thread interrupts upon the arrival of a new WC, to initiate two-sided synchronization.

The synchronization consists of a 2-way handshake between the two sides:
(1) Once encountering error (an error WC), side A resubmits its outstanding receive WQEs to the backup QP (QPs are bidirectional) and then sends a \textit{notification} to side B's backup QP, carrying side A's receive WQE counter and indicating that its receive WQEs are ready.
(2) Upon receiving this notification, side B resubmits its receive WQEs to its backup QP and sends an \textit{acknowledgment}, also carrying its receive WQE counter, back to side A's backup QP. Side B then resubmits its send WQEs to the backup QP.
(3) After receiving this acknowledgment, side A resubmits its send WQEs to the backup QP.
Detailed in \S\ref{sec:faulttolerance}.

\textbf{3. HOW to maintain execution order under active switching.} 
Maintaining WR execution order during active switching is challenging because RNICs execute WRs asynchronously with respect to software, which makes it difficult for SHIFT to determine the exact progress of WR execution on the hardware.

\textit{\underline{Solution: WR execution fence.}} We propose introducing a software fence for active switching. Specifically, SHIFT continues submitting unsignaled WRs to the backup QP until it submits a signaled WR, which serves as a \textit{fence WR}. After submitting the fence WR, SHIFT directs all subsequent WRs to the default QP but withholds the doorbell, ensuring that these WRs remain unexecuted. Only after SHIFT receives the WC for the fence WR does it update the doorbell on the default QP to begin execution. This mechanism ensures that the execution order surrounding the fence is strictly preserved.
Detailed in \S\ref{sec:recovery}.

\textbf{Architecture.}
Figure~\ref{fig:architecture} depicts the SHIFT architecture. For each QP, SHIFT implicitly creates a backup QP on a backup RNIC.
To tolerate anomalies that may occur anywhere along the path and to remain compatible with rail-optimized topologies (RNICs with the same index are connected to the same rail switch), SHIFT establishes backup QPs on the backup RNICs of both sides.

We implement SHIFT's core functions inside the control and data verbs of the RDMA userspace library~\cite{rdma-core}, called \emph{SHIFTLib}, while preserving verbs APIs for applications unchanged. Applications can adopt SHIFTLib by switching the RDMA library they use at runtime without rebuilding (e.g., updating \code{LD_LIBRARY_PATH} on Linux).

As illustrated in Figure~\ref{fig:architecture}, SHIFTLib modifies only the implementation of RDMA verbs while preserving the zero-copy data path (i.e., memory-RNIC-RNIC-memory). This design ensures that RDMA retains its low overhead and high performance.

Additionally, SHIFT deploys a KV storage over the management network to assist for establishing backup QPs.

\textbf{Execution flow.}
The logic of SHIFT involves two threads: an application thread that handles RDMA verb calls and a background thread responsible for managing backup RDMA resources and synchronizing two-sided operations. These threads operate across two phases: the RDMA setup phase and the data exchange phase.

\textit{RDMA setup phase.}
During this phase, the application thread initializes RDMA QPs using control verbs and captures these verbs along with their attributes. Simultaneously, the background thread executes the captured verbs to set up backup RDMA resources. (\S \ref{sec:setupphase})

\textit{Data exchange phase.} Once the RDMA connection is established, the application thread performs RDMA transfers using data verbs as usual, while the background thread waits for CQ events on backup QPs. (\S \ref{sec:defaultstate})

When a failure is detected on the sender\footnote{In RC mode, the sender ensures data delivery to receiver, and network failures manifest as error send WCs.}, the sender and receiver initiate a 2-way handshake to synchronize and fallback traffic to backup RNIC. (\S \ref{sec:faulttolerance})

While traffic uses backup RNICs, SHIFTLib periodically posts probe WRs to the default RNIC to assess its recovery status. Once the default RNIC has recovered, SHIFTLib switches the traffic back with the WR execution fence as described above. (\S \ref{sec:recovery})

\subsection{RDMA Setup Phase}
\label{sec:setupphase}
This section describes the RDMA setup phase of SHIFT.
SHIFT manages backup RNICs in an application-transparent and low-overhead manner.
Using a backup RNIC requires creating the corresponding backup RDMA resources (e.g., CQs and QPs). To avoid interference when RDMA resources are shared across processes, SHIFTLib allocates and tracks these backup resources on a per-process basis.

\begin{figure}[t]
    \centering
    \includegraphics[width=0.28\textwidth]{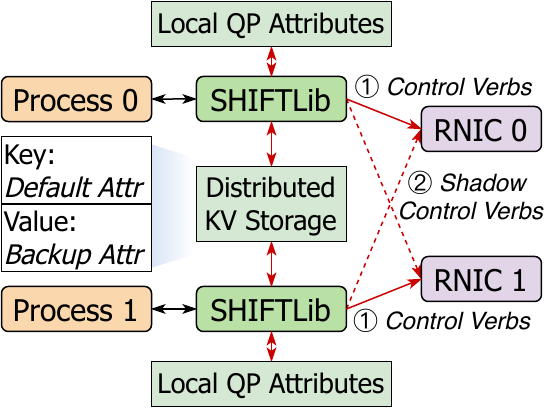}
    \setlength{\abovecaptionskip}{0pt}
    \setlength{\belowcaptionskip}{-10pt}
    \caption{
    \textit{SHIFT's control plane operations: When the application invokes control verbs, SHIFTLib records attributes, storing some locally and others in the KV store. A background thread in SHIFTLib then executes shadow control verbs using these recorded attributes.}
    }
    \label{fig:shadow_verbs}
\end{figure}

\begin{figure*}[t]
    \centering
    \includegraphics[width=0.9\textwidth]{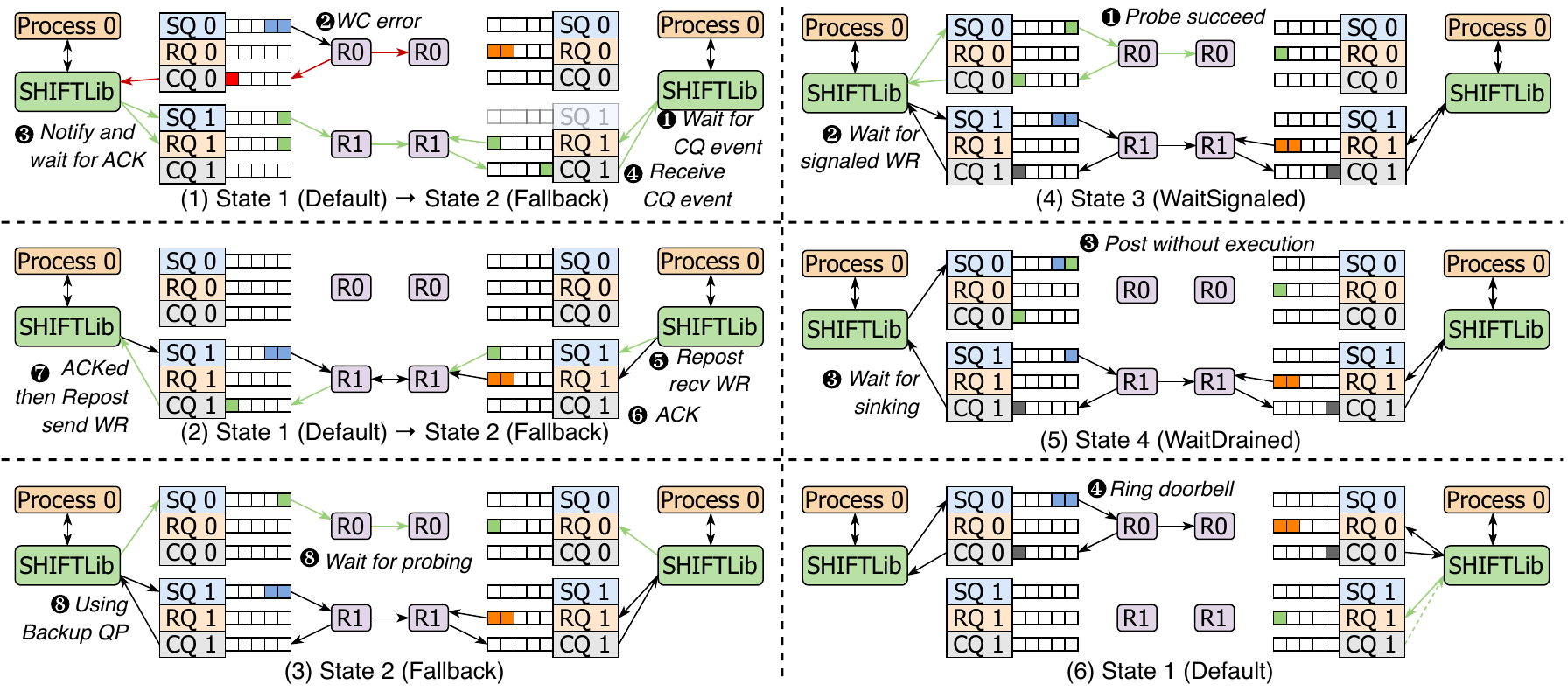}
    \setlength{\abovecaptionskip}{0pt}
    \setlength{\belowcaptionskip}{-10pt}
    \caption{
    \textit{Brief overview of the SHIFT state machine. The indices label the steps taken during the state transitions. Green arrows and boxes denote SHIFT control flow and WRs and WCs.}
    }
    \label{fig:controlflow}
\end{figure*}

\underline{\emph{Shadow control verbs.}} 
To manage backup RDMA resources without adding complexity to applications, SHIFT introduces \emph{shadow control verbs}. As illustrated in Figure~\ref{fig:shadow_verbs}, SHIFTLib implicitly initializes backup RDMA resources by replaying the same control verbs, with the same attributes, that the application used to create default resources. For example, when an application registers a memory region, SHIFTLib also registers the same region on the backup RNIC.
Shadow control verbs ensure that backup resources are ready before SHIFT falls back traffic to the backup RNIC.

To minimize the impact of additional control plane operations on the application, SHIFTLib executes shadow control verbs in background threads. 
When the application invokes control verbs to set up default RDMA resources, SHIFTLib records these verbs and their attributes. SHIFTLib concurrently runs a background thread that executes the recorded verbs on the backup RNICs. To avoid creating excessive threads, for each process, SHIFTLib assigns one control thread per RNIC to execute the shadow verbs for all QPs associated with that RNIC.

\underline{\emph{Out-of-band key-value (KV) transfer.}} 
An RDMA connection requires an out-of-band channel, typically TCP, to exchange QP route attributes (global ID (GID), QP number (QPN), and local ID (LID)) and memory region attributes (e.g., the remote memory region key (rkey)). Because SHIFT is application-transparent, it cannot assume access to the application's out-of-band channel. SHIFT therefore uses a key-value store over the management network to exchange backup attributes without interfering with the data plane.
The KV entries store \emph{<default memory region attributes $\rightarrow$ backup memory region attributes>} and \emph{<default QP route attributes $\rightarrow$ backup QP route attributes>}. Given a peer's default attributes, which is managed by the application, SHIFTLib can query the corresponding backup attributes.

Specifically, after initializing the local backup MR with \code{ibv_reg_mr} and backup QPs with \code{ibv_create_qp}, SHIFTLib puts these mappings to the KV store.

Before invoking \code{ibv_modify_qp} to configure a backup QP, SHIFTLib queries the KV store for the remote backup QP route attributes using the remote default route attributes. SHIFTLib similarly queries remote backup memory region attributes on demand, before their first use, which occurs when SHIFTLib first attempts to send on the backup RNIC. Retrieved attributes are cached to avoid redundant queries.

\underline{\emph{Overhead of backup QPs.}}
Although several previous studies have reported that excessive QPs may cause performance degradation due to on-chip QP context cache misses~\cite{srnic,1rma, understandingrnic}, this issue does not affect SHIFT. This is because the backup QPs are used only during network anomalies and remain idle under normal conditions. 
As a result, the backup QPs do not compete with the default QPs for on-chip cache and therefore do not introduce additional performance overhead. This is confirmed by the evaluation in \S\ref{sec:overhead}.

Moreover, although the backup QPs consume additional memory resources, this overhead is acceptable. When control verbs create a new QP, the QP context requires 3120 Bytes of memory, each SEND WQE requires 256 Bytes, and each RECV WQE requires 16 Bytes. Considering a send queue with 512 entries and a receive queue with 256 entries (the default value of NCCL), the total memory consumption for each backup QP is ~138 KB. Similarly, a CQ context requires 592 Bytes, and each CQE requires 64 Bytes. Each backup CQ with 512 entries consumes ~33 KB. Therefore, even if SHIFT creates 1k backup QPs on a server, the total memory consumption for backup QPs and CQs is only about 171 MB, which is negligible for modern servers.

\underline{\emph{Scalability of the key-value store.}} SHIFT utilizes a cluster-level key-value store to facilitate the initialization of backup RDMA resources. Common key-value stores, such as Redis~\cite{redis}, demonstrate the capacity to scale to thousands of clients and hundreds of thousands of requests per second~\cite{elasticache}, which far exceeds the requirements for SHIFT. Furthermore, because all interactions with the key-value store occur in a background thread, the latency and scalability of the store do not constitute a bottleneck for SHIFT.

\subsection{Data Exchange Phase: State Machine}
\label{sec:shiftstatemachine}
We design the \textit{SHIFT state machine} to manage RDMA traffic and resources for each default QP and its backup QP. It comprises four send-queue states:

\begin{enumerate}[noitemsep,parsep=0pt,partopsep=0pt,topsep=2pt, label=\textbf{State~\arabic*}, align=left, leftmargin=*]
    \item \textbf{(Default)}: The default QP operates normally with no failures currently detected. 
    \item \textbf{(Fallback)}: Traffic has switched to the backup QP while SHIFTLib waits for default QP to recover. 
    \item \textbf{(WaitSignaled)}: The default QP has recovered and is awaiting the next signaled WR from application. 
    \item \textbf{(WaitDrained)}: SHIFTLib awaits completion of the last signaled WR on the backup QP before switching traffic back to the default QP. 
\end{enumerate}

The receive queue has two states, \textbf{State 1 (Default)} and \textbf{State 2 (Fallback)}, indicating whether the default QP or backup QP handles incoming traffic.

We next describe the data exchange phase by walking through these send-queue states, as illustrated in Figure~\ref{fig:controlflow}.
\subsubsection{Default State}
\label{sec:defaultstate}
In State 1 (Default), the system runs normally with no detected failures. The default QP handles all RDMA traffic, while the backup QP remains idle. SHIFTLib maintains counters for the send and receive queues to track the number of completed two-sided send and receive WRs.

SHIFTLib posts a RECV WR to the backup QP and then asynchronously waits for its CQ event in a background thread (used by the handshake described in \S\ref{sec:designoverview}).

\subsubsection{Fault Tolerance (State 1\textrightarrow State 2)}
\label{sec:faulttolerance}
When network anomalies occur, including fatal failures or interface flapping, RDMA traffic times out and the RNIC places an error WC in the CQ. After polling this WC, SHIFT initiates the RDMA fallback process.
A complete fallback process handles traffic in both directions.

\underline{\emph{Retransmission-safe check.}} Before resubmitting send WQEs, SHIFTLib first verifies that all outstanding WQEs are safe for retransmission, following the boundary defined in \S\ref{sec:insight_boundary}. Specifically, SHIFTLib scans outstanding WQEs to detect atomic operations. If any are detected, SHIFT propagates the error to the application instead of attempting fallback. 

\underline{\emph{Work queue resubmitting.}} 
SHIFT resubmits WQEs by copying the WQE records in the send and receive work queues.

For send WQEs, SHIFTLib then compares the send and receive WQE counters to infer receive-side progress. If the receive-side progress is ahead, which indicates that the data arrived but the corresponding ack was lost, SHIFTLib treats the corresponding send WQEs as completed and excludes them from the outstanding set to avoid retransmissions.

SHIFTLib then copies these outstanding WQEs to the backup QP after updating their MR keys, whose mappings between default and backup QPs are stored in the KV store during MR registration, as well as their WQE indices and QP number to match the backup QP configuration. For send WQEs, SHIFTLib rings the doorbell of the backup QP to notify the RNIC of the new entries. For receive WQEs, SHIFTLib updates the receive doorbell record, which maintains the head index of the receive queue.

\underline{\emph{Overall RDMA fallback procedure.}}
As described in \S\ref{sec:designoverview}, SHIFT employs a 2-way handshake to synchronize the sender and receiver. Assuming side A first gets an error WC, the overall fallback procedure is as follows (illustrated in Figure~\ref{fig:controlflow}):

(1) Side A resubmits its outstanding receive WQEs to the backup QP and then sends a \code{WRITE_WITH_IMM} to side B's backup QP, using side A's receive counter as the immediate value. The receive state on side A transitions to \textbf{State 2 (Fallback)}. Side A then waits for the acknowledgment.

(2) Upon receiving this notification, side B receives a CQ event, which triggers the background thread. The background thread resubmits side B's receive WQEs to its backup QP and sends a \code{WRITE_WITH_IMM}, also carrying side B's receive counter, back to side A's backup QP as the ack. Side B then resubmits its send WQEs to the backup QP. The send and receive states on side B transition to \textbf{State 2 (Fallback)}.

(3) After receiving this acknowledgment, side A resubmits its send WQEs to the backup QP. The send state on side A transitions to \textbf{State 2 (Fallback)}.

Thereafter, RDMA traffic is handled transparently by the backup QP, maintaining application continuity.

To prepare for potential network recovery, SHIFTLib also resets the default QP. This reset is required to make the error-state QP, set by the RNIC when a WC error occurs, reusable. The QP reset procedure mirrors the initialization performed by the application during setup. SHIFTLib then posts a receive WR to the default QP for the same purpose as in \S\ref{sec:defaultstate}.

\subsubsection{Recovery (State 2\textrightarrow State 3\textrightarrow State 4\textrightarrow State 1)}
\label{sec:recovery}
Under network interface flapping, the default path may recover intermittently. SHIFT therefore continuously tests for recovery and, when possible, switches traffic back without violating WR ordering.

\underline{\emph{Probing for default path recovery.}}  
To test whether the default path has recovered, SHIFT uses a ping-like probe. SHIFT-Lib posts a WR that sends an empty write packet through the default path. The probe succeeds only if the WR completes, which implies that both the request and its acknowledgment were delivered. If the probe fails, SHIFTLib resets the default QP and continues probing periodically.

\underline{\emph{Seamless recovery with ordering.}}
If the probe succeeds, SHI-FTLib restores traffic to the default QP.
As discussed in \S\ref{sec:designoverview}, SHIFT uses a WR execution fence to ensure seamless recovery while preserving the order of WR execution.
SHIFT executes this recovery procedure independently in each direction.
Specifically, SHIFT switches traffic using the following steps:

\begin{enumerate}[noitemsep,parsep=0pt,partopsep=0pt,topsep=2pt,leftmargin=15pt]
    \item Once the probe succeeds, the SHIFT state machine transitions from State 2 (Fallback) to \textbf{State 3 (WaitSignaled)}. SHIFT continues posting WRs to, and polling the CQ from, the backup QP until the application posts the next signaled WR. This step is necessary because SHIFT can only rely on WCs, which are reported only for signaled WRs, to determine when all prior WRs have completed. We refer to this next signaled WR as the \emph{fence WR}.
    \item After the fence WR is posted to the backup QP, the state machine transitions to \textbf{State 4 (WaitDrained)}. SHIFTLib first posts a notification WR (i.e., a \code{WRITE_WITH_IMM} WR) to the default QP and enqueues all subsequent application WRs to the default QP. Importantly, SHIFT \textit{withholds the doorbell} of the default QP, which means these WRs will not be executed by the RNIC. This step ensures that the notification is processed before any later WRs after the switch.
    \item After all outstanding WRs on the backup QP have completed (as indicated by polling the corresponding WC of the fence WR), SHIFTLib first rings the doorbell only for the notification WR (the \code{WRITE_WITH_IMM} WR) on the default QP and waits for its completion. When the receiver observes this \code{WRITE_WITH_IMM}, a CQ event triggers its background thread, which runs the receive-side procedure in \S\ref{sec:faulttolerance} to repost outstanding receive WRs. 
    
    SHIFTLib then rings the doorbell for all WRs that were posted but not executed on the default QP. After that, the SHIFT state machine transitions back to \textbf{State 1 (Default)}, and subsequent WRs are posted to the default QP directly.
\end{enumerate}

\emph{We include more implementation details in Appendix~\ref{app:implementationdetails}}.

\subsection{Discussion}
\label{sec:discussion}

\textbf{RNICs with different attributes.} As described in \S\ref{sec:setupphase}, shadow control verbs initialize and manage backup RDMA resources by replaying the same verbs with the same attributes.
Consequently, shadow control verbs require the backup RNIC to support the same verbs and attributes as the default RNIC.
To the best of our knowledge, RNICs within a server are typically compatible at the verbs level for ease of maintenance. We leave support for heterogeneous RNICs with different attributes to future work.

\textbf{Hardware constraints of SHIFT.} 
If all RNICs on a server are connected to a single ToR switch, that ToR is a Single Point of Failure (SPOF) and its failure cannot be bypassed by SHIFT. SHIFT cannot overcome such hardware constraints.
In contrast, rail-optimized networks~\cite{rpingmesh, gao2021scaling, ncclall2all, superpod} connect different RNICs on a server to different ToR switches. 
This configuration allows SHIFT's RNIC fallback mechanism to bypass ToR failures, improving overall resilience.

\textbf{Multi-tenant scenario.}
In a multi-tenant environment, SHIFT may require additional security policies, including: (1) all backup RNICs must belong to the same tenant to ensure isolation, and (2) the KV store must be accessible only to that tenant to prevent data leakage.
The design for multi-tenancy is beyond the scope of this paper and is left for future work.

\textbf{Optimization with checkpointing-restart.}
To mitigate the performance impact of SHIFT fallback, checkpointing-restart mechanisms can be made aware of network anomalies. For example, after fallback completes, the application can checkpoint promptly instead of waiting for the next scheduled checkpoint. This approach reduces future progress loss and limits the time spent under degraded throughput.

\section{Evaluation}
\label{sec:evaluation}
\begin{figure*}[t]
    \centering
    \includegraphics[width=0.9\textwidth]{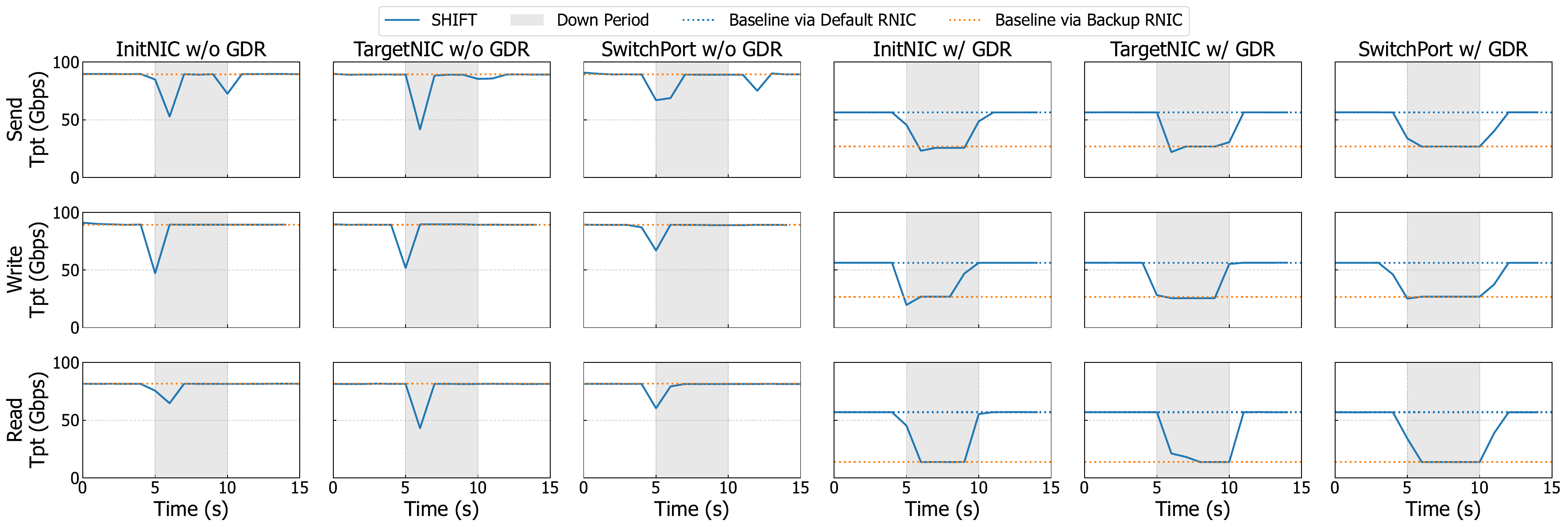}
    \setlength{\abovecaptionskip}{0pt}
    \setlength{\belowcaptionskip}{-10pt}
    \caption{
    \textit{Throughput measured using \texttt{ib\_send/write/read\_bw}~\cite{perftest}. Columns correspond to different failure injection methods, evaluated with and without GPUDirect RDMA (GDR). Failures are injected at $t=5$ s and recovered at $t=10$ s. 
    The one-second sampling interval causes the observed throughput drop to appear longer than the actual downtime (Fig~\ref{fig:micro-setup}(b)).
    GDR throughput is PCIe-bound due to our testbed's hardware topology (Fig~\ref{fig:micro-setup}(a)); however, this constraint does not affect the relative validity of the fallback mechanisms shown.
    Standard RDMA just terminates upon failure.
    }
    }
    \label{fig:tpt-time}
\end{figure*}

\begin{figure}[t]
    \centering
    \begin{subfigure}[b]{0.35\columnwidth}
        \centering
        \includegraphics[width=\linewidth]{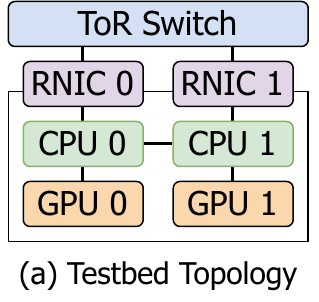}
    \label{fig:mbmtopology}
    \end{subfigure}
    \hfill
    \begin{subfigure}[b]{0.55\columnwidth}
        \centering
        \includegraphics[width=\linewidth]{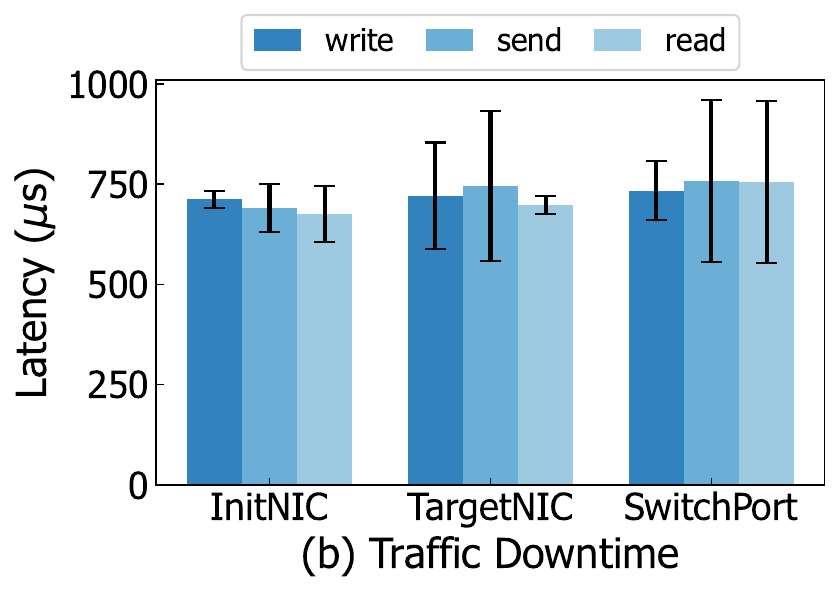}
        \label{fig:fallback-lat}
    \end{subfigure}
    \setlength{\abovecaptionskip}{0pt}
    \setlength{\belowcaptionskip}{-10pt}
    \caption{\textit{(a) Microbenchmark testbed topology. (b) Fallback latency of SHIFT under different failure scenarios.}}
    \label{fig:micro-setup}
\end{figure}

This section evaluates SHIFT using microbenchmarks and PyTorch-based distributed training. We demonstrate that SHIFT provides robust failure recovery with negligible performance overhead.

\noindent\textbf{Testbed.} Our testbed comprises two servers, each featuring an Intel Xeon Silver 4110 CPU (32 cores, 2.10 GHz), 128 GB RAM, and two ConnectX-5 100 Gb RNICs. The RNICs connect to a 100 GbE switch via copper cables and operate in RoCEv2 mode (MLNX\_OFED v5.8~\cite{ofed}). Additionally, each server hosts two NVIDIA P100 GPUs~\cite{p100}.

Due to hardware constraints of our testbed, RNICs and GPUs are connected via CPU and reside on different NUMA nodes, as depicted in Figure~\ref{fig:micro-setup}(a). Consequently, when GPUDirect RDMA (GDR)~\cite{gdr}, which enables direct GPU-RNIC data transfers, is enabled, throughput is bottlenecked by the QPI/PCIe interconnect rather than network.

\noindent\textbf{Implementation.} We implement SHIFTLib atop \texttt{rdma-core} v48~\cite{rdma-core}. The key-value store is implemented by Redis~\cite{redis} v8.0.5 over the control-plane network.

\noindent\textbf{Failure Injection.} We inject network anomalies using the following methods:
\begin{enumerate}[leftmargin=15pt] 
    \item \textbf{NIC failures.} We emulate RNIC faults by toggling the interface state via the operating system using standard Linux network utilities (e.g., \texttt{ip link set dev down/up}).
    \item \textbf{Switch/link failures.} We emulate physical link and switch failures by manipulating port states on the Top-of-Rack (ToR) switch via its management API. 
\end{enumerate}

\subsection{Microbenchmarks}
We evaluate SHIFT's ability to tolerate failures and its overhead with microbenchmarks in this section. 

\subsubsection{Handling Network Failures}
We evaluate SHIFT's fault tolerance using the \code{perftest} microbenchmark suite. We employ \code{ib_send_bw}, \code{ib_write_bw}, and \code{ib_read_bw} to generate two-sided (Send/Recv), one-sided RDMA Write, and one-sided RDMA Read traffic, respectively; these primitives constitute the majority of data transfers in distributed training.
We test three failure scenarios: initiator-side RNIC failure, responder-side RNIC failure, and switch port failure. For each scenario, we evaluate performance with GPUDirect RDMA (GDR)~\cite{gdr} enabled and disabled. Figure~\ref{fig:tpt-time} presents the throughput log results.

Additionally, we measure \textit{fallback latency}, defined as the interval between polling the first failed Work Completion (WC) and receiving the first successful WC after falling back to the backup RNIC (Figure~\ref{fig:micro-setup}(b)).
Note that perftest samples throughput every second, so the throughput fluctuations in Figure~\ref{fig:tpt-time} appear longer than the actual downtime.

The results demonstrate that:
(1) SHIFT incurs no throughput penalty for one-sided or two-sided operations in the absence of anomalies;
(2) SHIFT successfully sustains RDMA communication across all traffic types and failure scenarios by falling back to the backup RNIC;
(3) SHIFT seamlessly reverts traffic to the default RNIC upon device recovery.

\begin{figure}[t]
    \centering
    \includegraphics[width=0.4\textwidth]{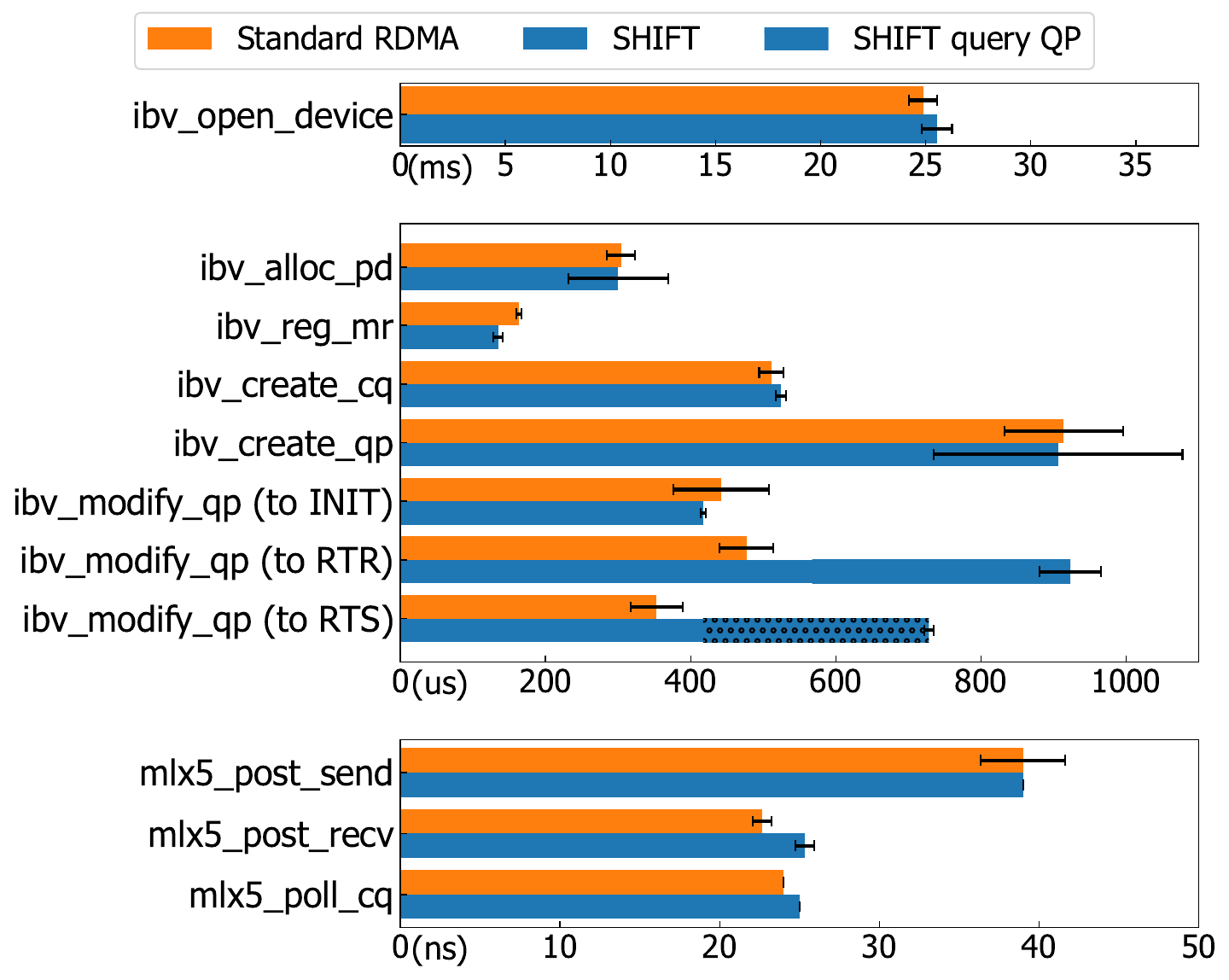}
    \setlength{\abovecaptionskip}{0pt}
    \setlength{\belowcaptionskip}{-10pt}
    \caption{
    \textit{Latency of data and control verbs in standard RDMA and SHIFT. Dotted blue bars indicate the overhead of \texttt{ibv\_query\_qp} during QP modification.}
    }
    \label{fig:verbs-lat}
\end{figure}

\begin{table}[t]\footnotesize
\centering
\begin{tabular}{lccc}
\toprule
\textbf{Bytes} & \textbf{Standard} & \textbf{SHIFT} & \textbf{Standard w/ 1000 QP} \\
\midrule
1 B & 2.68 ± 0.39 & 2.69 ± 0.38 & 2.69 ± 0.41 \\
2 B & 2.70 ± 0.44 & 2.70 ± 0.38 & 2.70 ± 0.39 \\
4 B & 2.70 ± 0.35 & 2.69 ± 0.37 & 2.71 ± 0.41 \\
8 B & 2.69 ± 0.40 & 2.71 ± 0.35 & 2.70 ± 0.43 \\
16 B & 2.69 ± 0.39 & 2.68 ± 0.38 & 2.70 ± 0.40 \\
\bottomrule
\end{tabular}
\setlength{\abovecaptionskip}{0pt}
\caption{\textit{Average latency (µs) ± std. dev. of \texttt{ib\_write\_lat} across message sizes.
}}
\label{tab:latency}
\end{table}

\begin{figure*}[t]
    \centering
    \begin{subfigure}[b]{0.68\textwidth}
        \centering
        \footnotesize
        \includegraphics[width=\linewidth]{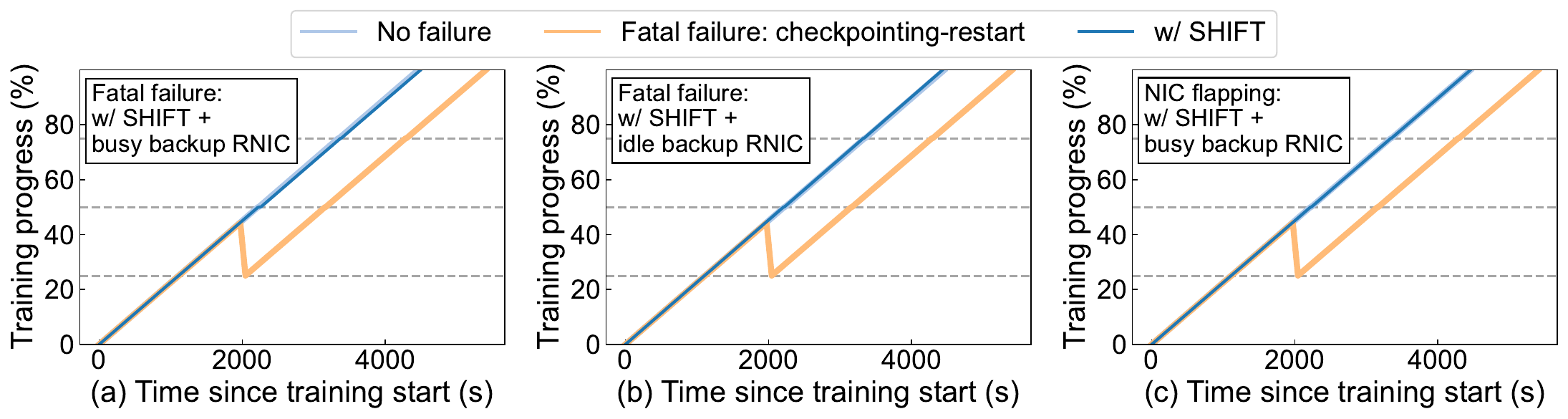}
    \end{subfigure}
    \begin{subtable}[b]{0.30\textwidth}
        \centering
        \footnotesize
        \begin{tabular}{lcccc}
            \toprule
            \textbf{Metric} & \textbf{(2)} & \textbf{(3)} & \textbf{(4)} & \textbf{(5)} \\
            \midrule
            Rescheduling time (s) & 63 & 37 & 0 & 0 \\
            Retrain from Ckpt (s) & 856 & 0 & 0 & 0 \\
            Total Slowdown (s) & 919 & 37 & 0 & 0 \\
            \bottomrule
        \end{tabular}
        \setcounter{subfigure}{3}
        \caption{{Detailed time of slowdown.}}
        \label{tab:slowdown-time}
    \end{subtable}
    \setlength{\abovecaptionskip}{0pt}
    \setlength{\belowcaptionskip}{-10pt}
    \caption{
    \textit{PyTorch training progress over time with and without SHIFT under different network conditions (\S\ref{sec:pytorch}). We take the no-failure case as an upper bound and checkpoint-restart without SHIFT as a baseline; dashed lines mark checkpoints. The three subfigures show that SHIFT reduces progress loss across scenarios. The table breaks down the slowdown components for settings (2)-(5).}
    }
    \label{fig:iters-time}
\end{figure*}

\subsubsection{Overhead of RDMA Verbs}
\label{sec:overhead}
To evaluate the performance overhead of SHIFT when \textit{no anomalies occur}, we measure the execution time of RDMA verbs and conduct an end-to-end latency microbenchmark using \code{ib_write_lat}.

\textbf{Execution Time of RDMA Verbs.} We measure the execution time of control verbs and data verbs when using the microbenchmark tool \code{ib_send_bw}. For control verbs, we evaluate the execution time of each verb in a single execution. For data verbs, we calculate the average execution time over 1,000,000 iterations for \code{post_send} and \code{post_recv}, and over 100,000,000 iterations for \code{poll_cq}. The execution times of RDMA verbs are shown in Figure~\ref{fig:verbs-lat}.

The results show that SHIFT introduces minimal overhead on the execution time of most control verbs, including \code{ibv_open_device}, \code{ibv_alloc_pd}, \code{ibv_reg_mr}, \code{ibv_create_cq}, \code{ibv_create_qp}, and \code{ibv_modify_qp} (to INIT). 
This indicates that storing control verbs and their attributes as shadow control verbs incurs negligible overhead.

Except for aforementioned control verbs, two control verbs experience greater overhead: the execution times for \code{ibv_modify_qp} (to RTR) and \code{ibv_modify_qp} (to RTS) are slower by 93.65\% and 106.34\%, respectively, because SHIFTLib queries the attributes of default QPs at these stages, which is essential for resetting default QPs after fallback to backup QPs (recall \S\ref{sec:faulttolerance}). 
The execution time of \code{ibv_query_qp} is marked in shadow in Figure~\ref{fig:verbs-lat}. 
The overhead of these control verbs is a one-time cost and thus remains acceptable for applications.

The execution time for shadow control verbs is about 35 ms, primarily due to \code{ibv_open_device} and KV transfer (recall \S\ref{sec:setupphase}). This overhead is acceptable because these verbs execute asynchronously without blocking the application and only during initialization.

The results for data verbs indicate that SHIFT introduces minimal overhead for \code{post_send}, \code{post_recv}, and \code{poll_cq} operations, as little additional processing is required when no anomalies occur.

\textbf{Write Latency.}
We employ \code{ib_write_lat}, a microbenchmark tool from the perftest package~\cite{perftest}, to measure the end-to-end write latency for message sizes of 1, 2, 4, 8, and 16 bytes in both standard RDMA and SHIFT. The comparison of end-to-end write latency between SHIFT and standard RDMA is shown in Table~\ref{tab:latency}.

The results indicate that SHIFT introduces near zero overhead on data path compared to standard RDMA in end-to-end write latency, with negligible impact on both average latency and standard deviation (Std).

\textbf{Overhead of Additional QP.} As discussed in \S\ref{sec:setupphase}, the additional backup QPs do not impact RDMA performance because they do not occupy RNIC cache when no anomaly occurs. We validate this claim with a simple experiment: we first create 1000 QPs on the RNIC without using them, which is sufficient to trigger the connection scalability issue reported in \cite{srnic}, and then rerun \code{ib_write_lat} to measure end-to-end latency. As shown in Table~\ref{tab:latency}, both the average latency and standard deviation remain nearly identical to standard RDMA without additional QPs.

\subsection{Real-World Application: PyTorch}
\label{sec:pytorch}
In this section, we evaluate SHIFT on PyTorch (v2.0.1)~\cite{pytorch}, a widely used framework for AI training, with NCCL (v2.19)~\cite{nccl}.

We train a GPT-2 model with 124M parameters~\cite{gpt2} using Distributed Data Parallel (DDP)~\cite{ddp} across two servers connected via RDMA. We use a global batch size of 8 on two servers, each equipped with two NVIDIA P100 GPUs. Given the limited GPU capacity, we restrict training to four epochs and configure the job to checkpoint once per epoch. To emulate network anomalies, we manually disable the RNIC on one server after 2000 seconds of training, which corresponds to approximately 1.78 epochs.

To demonstrate the benefits of SHIFT under network anomalies, we use the same dataset, model, hyperparameters, and failure injection, but vary the fault-tolerance mechanism and network conditions:
\begin{enumerate}[noitemsep,parsep=0pt,partopsep=0pt,topsep=2pt,leftmargin=15pt]
\item \textbf{No failure.} The training process runs without any network anomalies, and the observed training speed is used as the upper bound.
\item \textbf{Fatal failure: checkpointing-restart.} The training runs without SHIFT. After the job terminates, we restore the RNIC and rerun the training from the last checkpoint, simulating either anomaly resolution or task migration in production. This case serves as the baseline.
\item \textbf{Fatal failure: w/ SHIFT \& busy backup RNIC.} SHIFT is enabled while the backup RNIC carries other traffic (\code{ib_write_bw}). After the next checkpoint completes, the training terminates and resumes from that checkpoint.
\item \textbf{Fatal failure: w/ SHIFT \& idle backup RNIC.} The backup RNIC is idle, and after fallback to it, the training continues without performance interference.
\item \textbf{NIC flapping: w/ SHIFT \& busy backup RNIC.} The same busy backup RNIC as above is used, but the default RNIC is restored after 2 seconds.
\end{enumerate}

Figure~\ref{fig:iters-time}(a)-(c) plots training progress over time for these settings, and Figure~\ref{fig:iters-time}(d) reports the slowdown breakdown. In the baseline case, a network anomaly terminates the job, and checkpoint-restart avoids restarting from scratch but incurs rollback-induced progress loss. In our setup, rescheduling takes 63 seconds and retraining from the last checkpoint takes an additional 856 seconds.

Figure~\ref{fig:iters-time}(a) shows that with SHIFT, even when the backup RNIC is busy, training continues until the next checkpoint completes, eliminating rollback-induced progress loss. Training speed degrades minimally, suggesting that this workload is not network-bandwidth bound in our setup (given the relatively weak GPU). After the checkpoint completes, we gracefully stop the job; rescheduling takes 37 seconds, after which training proceeds normally.

Figure~\ref{fig:iters-time}(b) shows that if the backup RNIC is idle, training continues without measurable degradation after fallback. Figure~\ref{fig:iters-time}(c) shows that under NIC flapping, SHIFT switches RDMA traffic back to the default RNIC after recovery, leaving the end-to-end process nearly unaffected. In both cases, SHIFT prevents the training job from terminating.

These experiments demonstrate that SHIFT effectively mitigates training progress loss caused by network anomalies, while introducing negligible overhead in their absence.

\section{Related Works}
\label{sec:relatedworks}

\textbf{RDMA live migration.}
Software-based RDMA live migration aims to move an RDMA application to a new host while hiding changes in RNIC-maintained states; MigrRDMA achieves this by inserting an indirection layer in the RDMA driver, performing communication pre-setup during pre-copy, virtualizing RDMA identifiers via translation tables, and using wait-before-stop to preserve in-flight WR consistency~\cite{migrrdma}.
SHIFT targets uninterrupted training traffic on the same hosts, providing fault tolerance by falling back to other intra-host RNICs.

\textbf{Redesigned transports and alternative RMA abstractions.}
Recent datacenter transports relax or parameterize ordering to better exploit multipath and improve robustness (e.g., SRD’s reliable out-of-order delivery and Falcon’s ordered/weakly-ordered/unordered modes)~\cite{srd,falcon}.
1RMA further rethinks RMA as connection-free one-shot operations and delegates ordering and failure recovery to software when needed~\cite{1rma}.
SHIFT complements these efforts by keeping commodity verbs/RC and focusing on fast multi-RNIC failover and recovery for endpoint/interface faults rather than redefining the transport or programming model.

\textbf{Previous works on utilizing multiple RNICs.} Mellanox~\cite{mellanoxmprdma} introduced the idea of employing multiple RNICs for fault tolerance, but did not provide a detailed system design or evaluation. Their proposal also does not address recovery under interface flapping.
LubeRDMA~\cite{luberdma} was the first to present a detailed system design for RDMA traffic failover and recovery across multiple RNICs. In contrast, we make several improvements over LubeRDMA: we prove the RDMA failover trilemma (\S\ref{sec:insight_boundary}) and design a fault-tolerance mechanism with lower overhead, stronger support for two-sided operations and the recovery mechanism.

\section{Conclusion}
We presented SHIFT, a user-space RDMA layer that extends the ``fallback and bypass'' paradigm to the cross-NIC level for distributed training. We show a fundamental boundary: with commodity RNICs and a zero-copy data path, transparent failover cannot preserve RC guarantees. SHIFT exploits the fact that dominant training communication is largely idempotent and tolerant of relaxed memory ordering, enabling safe WQE-level retransmission across intra-host NICs. SHIFT adds negligible overhead and sustains execution through network anomalies, eliminating disruption to applications.

\clearpage

\bibliographystyle{ACM-Reference-Format}
\bibliography{bibliography}

\clearpage
\appendix
\section{RDMA Details}
\label{app:rdmadetail}
RDMA is widely adopted for high-throughput, low-latency communication in large-scale AI training. 
RDMA enables applications to transfer data between local and remote memory without CPU intervention. The RDMA workflow proceeds as follows: the application posts a work request (WR) for data transfer to the RNIC; the RNIC processes the WR and returns the result, known as a work completion (WC), to the completion queue (CQ); finally, the application polls the CQ for transfer results~\cite{rdma-core}.

RDMA typically offers three transport modes: Reliable Connection (RC), Unreliable Connection (UC), and Unreliable Datagram (UD)~\cite{transportmodes}. Among these, RC resembles TCP, whereas UD is datagram-based like UDP. UC effectively functions as a hybrid, offering unreliable connections similar to UDP but with connected semantics. 

To utilize RDMA, the application must allocate resources on both the sender and receiver sides using control verbs. It first opens a device (i.e., RNIC) for communication and obtains a context. Within this context, the application allocates several RDMA resources. First, it creates a Protection Domain (PD) to isolate resources. Then, it registers a Memory Region (MR) accessible by the RNIC, obtaining a local key (lkey) and a remote key (rkey) for subsequent local and remote access. Finally, it creates a Completion Queue (CQ) and a Queue Pair (QP). When creating the QP, the user specifies the local group ID (GID) index, which associates with the source IP and protocol version (e.g., RoCE v1 or v2). Notably, all allocated resources are unique to their contexts and isolated across RNICs.

Once resources are allocated, the application exchanges essential information, including QP details (e.g., QP number) and MR information (e.g., memory address and rkey). This exchange typically occurs via a TCP connection. Subsequently, the application modifies the QP to apply both local configuration and the received remote QP information. Upon modification, the QP transitions to Ready-to-Receive (RTR) or Ready-to-Send (RTS) state, enabling data transfer.

When a QP is ready, the application can initiate RDMA operations using various data verbs. For SEND/RECEIVE (two-sided) operations, the application must first post a receive WR to the receiver's queue and then a send WR to the sender's queue. Conversely, for READ/WRITE (one-sided) operations, receiver participation is unnecessary; the application simply posts a WR specifying the READ/WRITE operation to the send queue.

Send WRs, such as SEND or WRITE, trigger actual data transfers. These complete when the local RNIC receives an acknowledgment (ACK) from the remote RNIC confirming data written~\cite{qptype}. In contrast, receive WRs, such as RECEIVE, wait for incoming data transfers and complete once the data is successfully written to memory.

WR completions (i.e., WCs) are reported to the CQ associated with the QP, where the application retrieves them via polling. In RC mode, the RNIC executes WRs sequentially, ensuring that the completion order strictly matches both the execution and posting orders.

If a WR fails, the QP immediately enters an error state. Consequently, all posted but unexecuted WRs also fail, with their corresponding WCs reported to the CQ. While in this error state, any attempt to post new WRs results in an error. The QP must be transitioned back to the RTS state before processing further WRs.

\section{Implementation Details}
\label{app:implementationdetails}
\subsection{Avoiding deadlock control dependencies} 
Because SHIFTLib executes all shadow control verbs belonging to the same RNIC within a single background thread, deadlocks may occur when an application uses multiple threads to initialize default QPs. Suppose servers A and B each initialize QP 1 and QP 2 through control verbs in separate threads, and SHIFTLib records the executed verbs in a list as shadow control verbs. The order of these verbs in the list is non-deterministic. Since QP initialization involves creating a QP and then configuring it with the remote QP's attributes, it is evident that configuring a QP depends on the creation of its corresponding remote QP. A potential deadlock may arise during the execution of shadow control verbs in the background thread if server A has only completed creating QP 1 and querying server B QP 1's attributes, while server B has only completed creating QP 2 and querying server A QP 2's attributes. If the shadow control verbs are then executed strictly one by one, cyclic dependencies will emerge.

To break the cyclic dependencies, SHIFTLib executes shadow control verbs in a \textit{best-effort} manner. 
When executing a shadow control verb, if it depends on a remote attribute that is not ready yet, SHIFTLib skips this verb and tries to execute the next verb in the list.
SHIFTLib repeatedly scans the list until all verbs are executed.

\subsection{WC buffer}
When preparing to resubmit outstanding WQEs to the backup QP, SHIFTLib first needs to determine the outstanding WQEs. SHIFTLib polls the default CQ to obtain successful but unpolled WCs, storing them in a local WC buffer~\footnote{These WCs will later be passed to the application when it polls the CQ}, and continues until it observes the first error WC. The remaining WQEs in the send and receive queues are then identified as outstanding.

\section{Formal Verification of Failover Impossibility}
\label{app:proofs}

This appendix provides the exact correspondence between our Rocq definitions/theorems and their textual descriptions. We model the system using a small step operational semantics. All proofs are mechanically verified in Rocq 9.0 ($\sim$3,900 lines) without additional axioms.

\subsection{Core Definitions}

\textbf{Memory Model.}
We model memory addresses and values as natural numbers.
\begin{itemize}
    \item \textbf{Address}: An address $a \in \mathbb{N}$ identifies a location in RDMA-accessible memory.
    \item \textbf{Value}: A value $v \in \mathbb{N}$ represents data stored at a memory address.
    \item \textbf{Memory}: A function $m : \text{Addr} \to \text{Val} $ mapping each address to its current value.
    \item \textbf{Initial Memory}: The initial memory $m_0$ satisfies $m_0(a) = 0$ for all addresses $a$.
\end{itemize}
Operations on memory are defined as:
\begin{itemize}
    \item \texttt{mem\_read(m, a)}: Returns $m(a)$.
    \item \texttt{mem\_write(m, a, v)}: Produces memory $m'$ where $m'(a) = v$ and $m'(a') = m(a')$ for $a' \neq a$.
\end{itemize}

\textbf{RDMA Operations.}
An RDMA operation is one of:
\begin{itemize}
    \item $\text{Write}(a, v)$: Write value $v$ to address $a$.
    \item $\text{Read}(a)$: Read value at address $a$.
    \item $\text{FADD}(a, \delta)$: Atomically add $\delta$ to address $a$, return old value.
    \item $\text{CAS}(a, \text{exp}, \text{new})$: If $m(a) = \text{exp}$, write $\text{new}$ and return (success, old value). Else return (failure, old value).
\end{itemize}

Execution semantics \texttt{exec\_op} are defined in the standard way. For example, $\text{exec\_FADD}(m, a, \delta)$ atomically reads $v_{old} = m(a)$, writes $m(a) := v_{old} + \delta$, and returns $v_{old}$.

\textbf{Execution Traces.}
A trace $\mathcal{T}$ is a list of events representing one possible distributed execution. Events include:
\begin{itemize}
    \item \textbf{Sender-side}: \texttt{EvSend(op)}, \texttt{EvCompletion(op, res)}, \texttt{EvTimeout(op)}.
    \item \textbf{Network}: \texttt{EvPacketLost(op)}, \texttt{EvAckLost(op)}.
    \item \textbf{Receiver-side}: \texttt{EvReceive(op)}, \texttt{EvExecute(op, res)}.
    \item \textbf{Application}: \texttt{EvAppConsume(a, v)}, \texttt{EvAppReuse(a, v)}.
\end{itemize}

\textbf{Sender View.}
The sender view $\sigma(\mathcal{T})$ projects a trace to only sender-observable events (Send, Completion, Timeout). This is the central abstraction: the sender cannot observe network losses or receiver execution events directly.
Two traces $\mathcal{T}_1, \mathcal{T}_2$ are \emph{sender-indistinguishable} if $\sigma(\mathcal{T}_1) = \sigma(\mathcal{T}_2)$.

\subsection{Proof of Lemma~\ref{lem:impossibility} (Indistinguishability)}
\label{app:proof_indistinguishability}

We identify two concrete traces that are indistinguishable to the sender but require opposite actions.
Let $W_D(v)$ be an RDMA Write of value $v$ to data address $A_{data}$.

\textbf{Trace 1: Packet Loss ($\mathcal{T}_1$).}
The sender posts $W_D(V_1)$, but the packet is lost in the network (\texttt{EvPacketLost}). The sender eventually times out (\texttt{EvTimeout}). In this trace, the operation was \textbf{not executed}. To ensure liveness, the correct action is to \textbf{retransmit}.

\textbf{Trace 2: ACK Loss + Reuse ($\mathcal{T}_2$).}
The sender posts $W_D(V_1)$, and it is successfully executed by the receiver (\texttt{EvExecute}). The application then consumes the data and reuses the buffer for a new value $V_{new}$ (\texttt{EvAppReuse}). However, the acknowledgment is lost (\texttt{EvAckLost}), causing the sender to timeout (\texttt{EvTimeout}). In this trace, the operation \textbf{was executed}. To ensure safety (avoiding the overwrite of $V_{new}$), the correct action is to \textbf{do not retransmit}.

\begin{theorem}[Indistinguishability]
$\sigma(\mathcal{T}_1) = \sigma(\mathcal{T}_2)$.
\end{theorem}
\textit{Proof.} In both traces, the sender observes identical events: the initial post (\texttt{EvSend}) followed by a timeout (\texttt{EvTimeout}). The intermediate events (network loss vs. execution and reusing) are hidden from the sender. Consequently, the sender view $\sigma(\mathcal{T})$ is identical for both $\mathcal{T}_1$ and $\mathcal{T}_2$. 

Since the overlay depends only on the sender view, the overlay will return identical transmit decisions, either both retransmit, or both do not retransmit, violating either liveness or safety. In general, no deterministic function depending only on $\sigma(\mathcal{T})$ can correctly distinguish between the need to retransmit (Trace 1) and the need to abort (Trace 2).

\subsection{Proof of Lemma~\ref{lem:non-idempotency} (Non-Idempotency)}
\label{app:proof_non_idempotent}

\begin{lemma}[FADD Non-Idempotency]
For any $\delta \neq 0$, $\text{FADD}(a, \delta)$ is not idempotent. 
See Rocq formalization \texttt{fadd\_non}
\texttt{\_idempotent}.
\end{lemma}

\begin{lemma}[CAS Double Success / ABA]
Under concurrent modification, a CAS retry can succeed even after the original succeeded. See Rocq formalization \texttt{cas\_double\_success}.
\end{lemma}

These results of non-idempotency of atomic operations are long standing results. We extend the non-idempotency results to include two-sided operations and modern high-performance communication patterns like those found in collective communication libraries.

\begin{lemma}[Two-Sided Consumption]
An RDMA Send operation is non-idempotent with respect to the receiver's Work Queue. See Rocq formalization \texttt{send\_non\_idempotent}.
\end{lemma}

\textit{Proof.} Let the receiver state be defined by a sequence of posted receive buffers $Q = [b_1, b_2, \dots, b_n]$.
\begin{enumerate}
\item \textbf{Execution}: $\text{EvExecute}(\text{Send}(v))$ consumes the head of the queue, resulting in $Q' = [b_2, \dots, b_n]$ and $m(b_1) = v$.
\item \textbf{Indistinguishability}: In Trace $\mathcal{T}_2$, the ACK is lost. The sender view $\sigma(\mathcal{T}_2)$ contains only a timeout, which is identical to the packet loss scenario $\sigma(\mathcal{T}_1)$.
\item \textbf{Retry}: The sender retries $\text{Send}(v)$. The receiver hardware, having no record of the previous execution due to the lost ACK state, matches this request against the next available buffer $b_2$.
\item \textbf{Violation}: The final state is $Q'' = [b_3, \dots, b_n]$. A single logical send has consumed two physical buffers and corrupted the memory location $b_2$, which was intended for a subsequent application message.
\end{enumerate}

\begin{lemma}[Write-after-Reuse / NCCL LL]
For protocols that pack data and control flags in a single atomic write (e.g., NCCL LL/LL128), a retry after a lost ACK violates memory safety under application reuse.
\end{lemma}

\textit{Proof.} Let $m(a)$ be the memory at address $a$, where $a$ stores a tuple $(v, f)$ representing data and a sequence flag.
\begin{enumerate}
\item \textbf{Initial State}: $m(a) = (V_0, F_0)$.
\item \textbf{Operation}: Sender issues $W(a, (V_1, F_1))$. Receiver executes this via $\text{EvExecute}$. The application, polling on address $a$, observes the updated flag $F_1$ and consumes data $V_1$.
\item \textbf{Reuse}: Via $\text{EvAppReuse}$, the application reallocates address $a$ for a new local computation, resulting in $m(a) = (V_{new}, F_{new})$.
\item \textbf{ACK Loss}: The original ACK is lost ($\text{EvAckLost}$). The sender observes a timeout and, following a deterministic retransmission policy, retries $W(a, (V_1, F_1))$.
\item \textbf{Corruption}: The retry overwrites $m(a)$ with the stale $(V_1, F_1)$. Because $F_1$ may still be numerically valid (e.g., in a circular buffer where flags are reused), the application reads $V_1$ as if it were the most recent update, leading to silent data corruption (SDC).
\end{enumerate}

\subsection{Proof of Theorem~\ref{thm:consensus} (Consensus Hierarchy Barrier)}
\label{app:proof_consensus}

We prove that constructing a transparent failover mechanism that ensures both safety and liveness is impossible on standard RDMA hardware.

\textbf{Definition (Consensus Hierarchy).}
The Consensus Hierarchy, introduced by Herlihy~\cite{herlihy1991waitfree}, classifies synchronization primitives based on their ability to solve the wait-free consensus problem for $n$ processes.
\begin{itemize}
    \item \textbf{Consensus Number 1 (CN=1)}: Primitives that cannot even solve consensus for 2 processes (e.g., Read/Write).
    \item \textbf{Consensus Number 2 (CN=2)}: Primitives that can solve consensus for 2 processes (e.g., Test-and-Set, Fetch-and-Add).
    \item \textbf{Consensus Number $\infty$ (CN=$\infty$)}: Primitives that can solve consensus for any number of processes (e.g., CAS).
\end{itemize}

\begin{theorem}[Consensus Hierarchy Barrier]
Transparent failover requires a system with $\text{CN} \ge 2$, but transparent verification on standard RDMA is restricted to $\text{CN} = 1$.
\end{theorem}
\textit{Proof.} We proceed by reduction, showing that a correct failover mechanism implies the ability to solve 2-process consensus.

\textbf{1. Abstract Interfaces.}
We define two abstract objects:
\begin{itemize}
    \item \textbf{Sticky Register ($\mathcal{S}$)}: A shared register that enforces a "First-Writer-Wins" policy. The first value written is permanently committed; subsequent writes are ignored.
    \item \textbf{Failover Mechanism ($\mathcal{F}$)}: A system that manages two entities: a "Ghost" (state pending in the network) and a "Backup" (recovery attempt). Safety requires that if the Ghost executes first, the Backup must abort. Liveness/Stability requires that if the Backup recovers first, it must block the Ghost.
\end{itemize}

\textbf{2. The Reduction ($\mathcal{F} \implies \mathcal{S}$).}
We can construct a Sticky Register using a correct Failover Mechanism. We map the race for the register to the race between the Ghost and Backup:
\begin{itemize}
    \item \textbf{Process A writes to $\mathcal{S}$}: Modeled as the Ghost packet arriving.
    \item \textbf{Process B writes to $\mathcal{S}$}: Modeled as the Backup initiating recovery.
\end{itemize}
The Failover Mechanism $\mathcal{F}$ guarantees that exactly one of these actions "wins" (takes effect) while the other is rejected or has no effect. This exactly matches the semantics of a Sticky Register.

\textbf{3. Sticky Register solves 2-Consensus.}
A Sticky Register $\mathcal{S}$ is powerful enough to solve consensus for 2 processes. Both processes propose a value by attempting to write to $\mathcal{S}$. Afterwards, they read $\mathcal{S}$. Since $\mathcal{S}$ is "sticky", both will read the same value (the value of the first successful write). Thus, they reach agreement.
This implies $\text{CN}(\mathcal{S}) \ge 2$. By the reduction, $\text{CN}(\mathcal{F}) \ge 2$.

\textbf{4. Impossibility under Non-responsive Failures.}
In RDMA, packet or ACK loss manifests as a timeout where the sender cannot distinguish whether the operation succeeded or not. This precisely matches the definition of \textbf{non-responsive failures} (specifically, NR-Omission) introduced by Jayanti et al.~\cite{jayanti1998fault}, where a base object may ``hang'' without responding (timeout) while other process may see its completed result.

They have shown that \emph{even the most benign non-responsive failure mode cannot be tolerated} when constructing a wait-free shared object with Consensus Number $N \ge 2$ from base objects subject to such failures, unless randomization is used~\cite{jayanti1998fault}.

As established in Step 3, safe failover requires a \textbf{Sticky Register}, which has $\text{CN}=2$. However, our base objects (commodity RNICs) are subject to non-responsive failures. Therefore, a deterministic, wait-free implementation of a Sticky Register is fundamentally impossible on standard RDMA.

\textbf{Conclusion.}
We have shown that any correct Failover Mechanism requires synchronization power equivalent to at least Consensus Number 2 (a Sticky Register). However, standard RDMA primitives suffer from non-responsive failures (timeouts), which prevents the deterministic construction of any object with $\text{CN} \ge 2$~\cite{jayanti1998fault}. Thus, reliable, deterministic cross-NIC failover is impossible without stronger hardware primitives or randomization.

\end{document}